\documentclass[twocolumn,nopacs]{revtex4}

\usepackage{amsmath,amssymb,graphicx}

\pdfoutput=1
\usepackage{hyperref}
\hypersetup{pdfauthor={KSD Beach and FF Assaad},pdftitle={Bilayer Hubbard model for 3He: a cluster dynamical mean-field calculation}}

\newcommand{\bracket}[1]{\langle #1 \rangle}
\renewcommand{\vec}[1]{\mathbf{#1}}

\DeclareMathOperator{\Imag}{Im}
\DeclareMathOperator{\Tr}{Tr}

\begin{document}

\title{Bilayer Hubbard model for $^3$He: a cluster dynamical mean-field calculation}
\author{K. S. D. Beach$^{(1)}$ and F. F. Assaad$^{(2)}$}

\affiliation{$^{(1)}$Department of Physics, University of Alberta, Edmonton, Alberta, Canada T6G 2G7 \\
$^{(2)}$ Institut f\"ur theoretische Physik und Astrophysik,
Universit\"at W\"urzburg, Am Hubland, D-97074 W\"urzburg, Germany}

\begin{abstract}
Inspired by recent experiments on bilayer $^3$He, we consider a bilayer Hubbard model on a triangular lattice. 
For appropriate model parameters, we observe a band-selective Mott transition at a critical chemical potential, 
$\mu_c$, corresponding to the solidification of the fermions in the first layer. The growth of the effective mass
on the metallic side ($\mu < \mu_c$) is cut off by a first order transition in which the first layer fermions 
drop out of the Luttinger volume and their spin degrees of freedom become locked in a spin singlet state. 
These results are obtained from a cluster dynamical mean-field calculation on an eight-site cluster 
with a quantum Monte Carlo cluster solver. 
\end{abstract}


\maketitle
The solidification of $^3$He monolayers~\cite{Casey03} has been 
interpreted as a density-driven Mott transition in which the effective 
mass diverges~\cite{Vollhardt_rev,Imada_rev}. Below 
the critical density, the metallic phase corresponds to a nearly localized Fermi liquid and, 
beyond the critical density, to a spin-disordered solid. 
Recently, it has been possible to realize bilayers of $^3$He~\cite{Neumann07} 
(atop a frozen $^4$He substrate, itself adsorbed onto graphite) 
with the special property that the second layer begins to form before the first has solidified. 
Since the first layer is close to a Mott transition, the $^3$He fermions in 
this layer are slow (i.e.\ heavy), whereas those in the second layer are fast. 
This combination of fast and slow dynamics---corresponding to wide and narrow fermion conduction
bands---is completely analogous to the situation in electronic heavy fermion materials, albeit
without the complication of crystal field and spin orbit effects.

According to this picture, prior to solidification of the first layer, one expects an 
enhanced effective mass and a Luttinger volume that counts both the first- and 
second-layer populations. Moreover, one naively anticipates that further $^3$He
deposition will eventually cause the effective mass to diverge. 
In experiment, the effective mass is indeed observed to increase as a function of the 
total $^3$He concentration, but its growth is interrupted by an intervening phase~\cite{Neumann07} 
in which the first $^3$He layer is a spin-disordered insulator, decoupled from the second layer.

\begin{figure}
\includegraphics{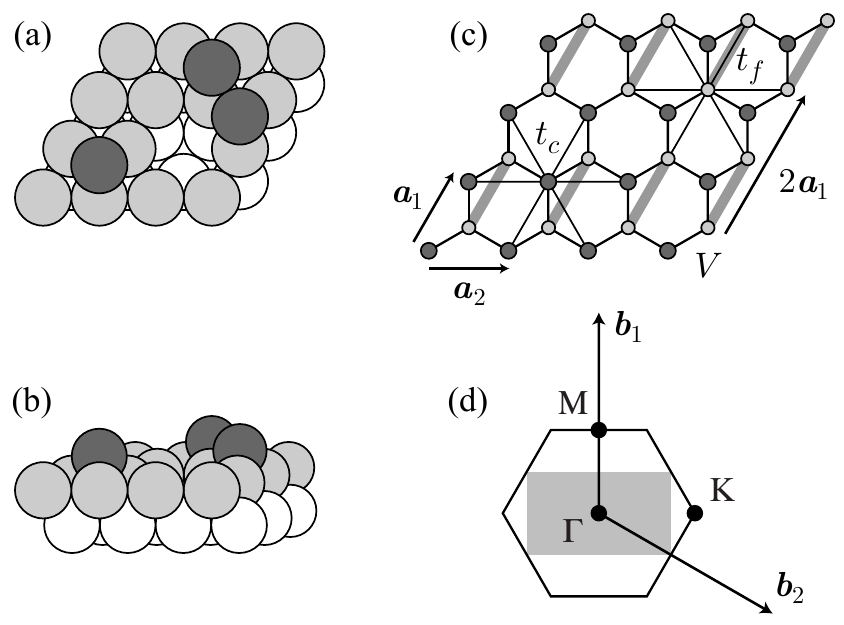} 
\caption{\label{Model.fig}
(a),(b) Stacking of billiard balls modeling of bilayer $^3$He, top and side view, with 
the $^4$He substrate shown in white.
(c) Tight-binding modeling with hoppings $t_c$, $t_f$, and $V$.
The grey bars connecting $f$ fermion sites (first-layer $^3$He positions) indicate
a possible singlet pattern compatible with the eight-site $2\vec{a}_1 \times 2\vec{a}_2$
and four-site $2\vec{a}_1 \times \vec{a}_2$ unit cells.
(d) The hexagonal Brillouin zone of the triangular lattice and the rectangular zone
that results from the folding $\vec{b}_1 \to \tfrac{1}{2}\vec{b}_1$.}
\end{figure}
 
The motivation for this Letter is to consider a simple lattice model that goes a 
good way toward reproducing the aforementioned experimental situation. 
As shown in Figs.~\ref{Model.fig}(a) and \ref{Model.fig}(b), we adopt a {\it stacking of billiard balls} modeling of 
bilayer $^3$He on a triangular lattice defined 
by $\vec{a}_1 =(1/2, \sqrt{3}/2,0)$ and $\vec{a}_2 = (1,0,0)$. Each unit cell 
accounts for two $^3$He positions, $\vec{r}_f = \vec{0}$ and $\vec{r}_c = 
\frac{2}{3} \vec{a}_1 - \frac{1}{3} \vec{a}_2 + (0,0,a_3)$, measured relative to the lattice.
Note that this presupposes a particular stacking arrangement for the second $^3$He layer.

Our model can be viewed as a honeycomb lattice whose inequivalent sites (corresponding
to $^3$He positions in the upper and lower layers) are populated by two species of fermion, 
which we label $c$ and $f$.
The tight-binding parameters include a nearest-neighbour (interlayer) hopping $V$ and next-nearest-neighbour (intralayer) hoppings $t_c$ and $t_f$. See Fig.~\ref{Model.fig}(c). With the inclusion of onsite Coulomb repulsion terms, the
Hamiltonian reads
\begin{equation} 
\begin{gathered}
 H = \sum_{\vec{k},\sigma} 
 \begin{pmatrix} c^{\dagger}_{\vec{k},\sigma} & f^{\dagger}_{\vec{k},\sigma} \end{pmatrix}
 \begin{pmatrix}
 \varepsilon_c(\vec{k}) - \mu & V(\vec{k}) \\
 V(\vec{k}) & \varepsilon_f(\vec{k}) - \mu
 \end{pmatrix}
 \begin{pmatrix}
 c_{\vec{k},\sigma} \\ f_{\vec{k},\sigma}
 \end{pmatrix} \\
{} + U_c \sum_{\vec{i}} \bigl( \hat{n}_{c,\vec{i}} - 1 \bigr)^2 
 + U_f \sum_{\vec{i}}\bigl( \hat{n}_{f,\vec{i}} - 1 \bigr)^2. 
\end{gathered}
\end{equation} 
Here, the mixing element $V(\vec{k}) = V(1+\gamma_{\vec{k}})^{1/2}$ and the
dispersion $\varepsilon_c(\vec{k}) = -t_c \gamma_{\vec{k}} + \varepsilon_c^{0}$
are expressed
in terms of the connection
$\gamma_{\vec{k}} = \cos(\vec{k}\cdot \vec{a}_1) + \cos(\vec{k}\cdot \vec{a}_2) + 
\cos[\vec{k}\cdot (\vec{a}_2-\vec{a}_1)]$
of the underlying Bravais lattice. The operator
$\hat{n}_{c,\vec{i}} = \sum_{\sigma} c^{\dagger}_{\vec{i},\sigma}
c_{\vec{i},\sigma}$ is the local $^3$He density in the upper layer.
Similar definitions hold for $\varepsilon_f(\vec{k})$ and $\hat{n}_{f,\vec{i}}$.

Except for the complication of the layer stacking (and the resulting $\vec{k}$-dependent
hybridization), this bilayer Hubbard model reduces to the Periodic Anderson Model 
as $t_f \rightarrow 0$, a limit in which the bare mass of the $f$ fermions diverges. 
Similar models (though ones with a somewhat unrealistic \emph{direct} layer stacking)
have been considered for the description of bilayer $^3$He in Refs.~\onlinecite{Benlagra08}
and \onlinecite{Benlagra09}
within a slave boson mean-field calculation. Here we go one step further and 
perform calculations within the cellular dynamical mean field theory (CDMFT)~\cite{Biroli04} approximation 
using a supercell defined by the lattice constants $L_c \vec{a}_1$ and $L_c \vec{a}_2$ with $L_c = 2$. 
Since the original unit cell contains two orbitals, this amounts to a total of eight orbitals
per supercell. The resulting single particle Green function,
 $\underline{G} ( \vec{K}, i \omega_m ) $, is a $2 L_c^2 \times 2 L_c^2$ matrix 
with crystal momentum $\vec{K}$ in the Brillouin zone of the supercell lattice.
The CDMFT calculation involves neglecting momentum conservation and 
thereby obtaining a $\vec{K}$-independent self-energy 
$\underline{\Sigma}(i \omega_m) $. This quantity is extracted from an $L_c \times L_c$ cluster of 
unit cells embedded in a dynamical mean field that is determined self-consistently. We have
solved this cluster problem using a standard Hirsch-Fye approach and have symmetrized the cluster 
Green function to obtain the corresponding quantity on the lattice:
\begin{equation}
\label{Greenk.eq}
 G(\vec{k},i\omega_m)_{\mu,\nu} = \frac{1}{L_c^2} \sum_{\alpha,\beta} e^{i \vec{k} \cdot
 \left( \vec{x}_{\alpha} - \vec{x}_{\beta} \right)} 
 \underline{G} ( \vec{K}, i \omega_m )_{(\mu,\alpha), (\nu, \beta)}. 
\end{equation}
Here $\vec{x}_{\alpha}$ denotes the unit cell positions within the supercell, 
$\mu$ and $\nu$ run over the $c$ and $f$ orbitals within each unit cell, 
and $\vec{k}$ and $\vec{K}$ differ by a reciprocal lattice vector of the supercell 
Bravais lattice. The rotation to real frequencies was carried out with 
a stochastic analytical continuation technique~\cite{Sandvik98,Beach04a}. 

\begin{figure}
\includegraphics{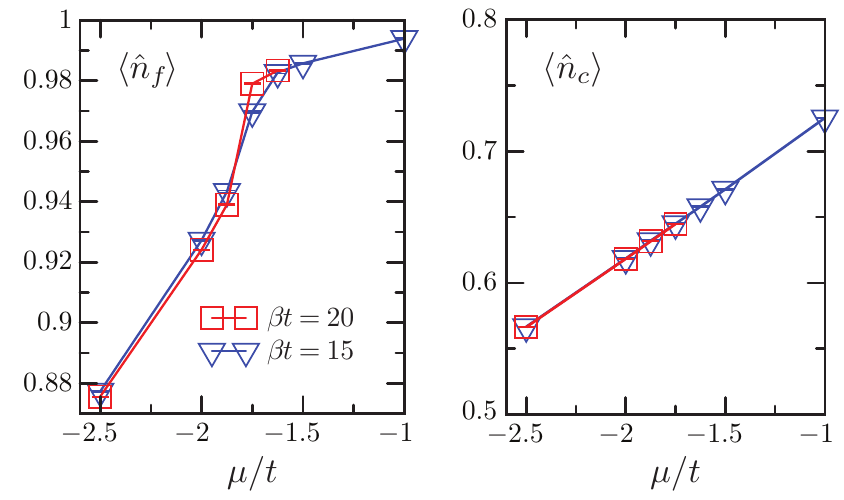}
\caption{\label{Density.fig}
Single particle occupation number for the $f$ and $c$ fermions as a function of temperature.} 
\end{figure}

We consider the following model parameters: $t_c = t_f = t$, $U_c/t = U_f/t = 12$, 
$V_0/t = 1/2$, $\varepsilon^0_c/t = 3$ and $\varepsilon^0_f/t = 0$. We have chosen large values 
of $U_c$ and $U_f$ to reflect the contact repulsion of the $^3$He atoms and to guarantee 
that each single layer is well within the Mott insulating phase at half-band filling~\cite{Kyung07}. 
These values of the Hubbard interaction lead to low double occupancy,
thus generating local moments. The difference $\varepsilon^0_c - \varepsilon^0_f > 0$ is a
crude accounting for the van der Waals forces (both $^4$He--$^3$He and $^3$He--$^3$He)
that preferentially fill the first layer. Fig.~\ref{Density.fig} plots the layer
densities $\bracket{\hat{n}_f}$ and $\bracket{\hat{n}_c}$ as a function of the chemical potential,
which controls the overall $^3$He concentration.
Analysis of the temperature dependence of $\bracket{\hat{n}_f}$ is consistent with 
a zero-temperature jump of this quantity at $\mu = \mu_c \simeq -1.8 t $. In 
contrast, $\bracket{\hat{n}_c}$ increases smoothly with the chemical potential.
Since $\frac{\partial F}{\partial \mu} = \bracket{\hat{n}_c + \hat{n}_f}$, the jump in the 
total fermionic density signals a density-driven first order transition. 
In a canonical ensemble, states with total density lying within the jump are phase separated.
The first order nature of this phase transition can be confirmed explicitly on 
smaller four-site clusters, which can be simulated at much lower temperatures before the negative sign 
problem becomes unmanageable~\cite{companion_paper}.

\begin{figure}[b]
\includegraphics{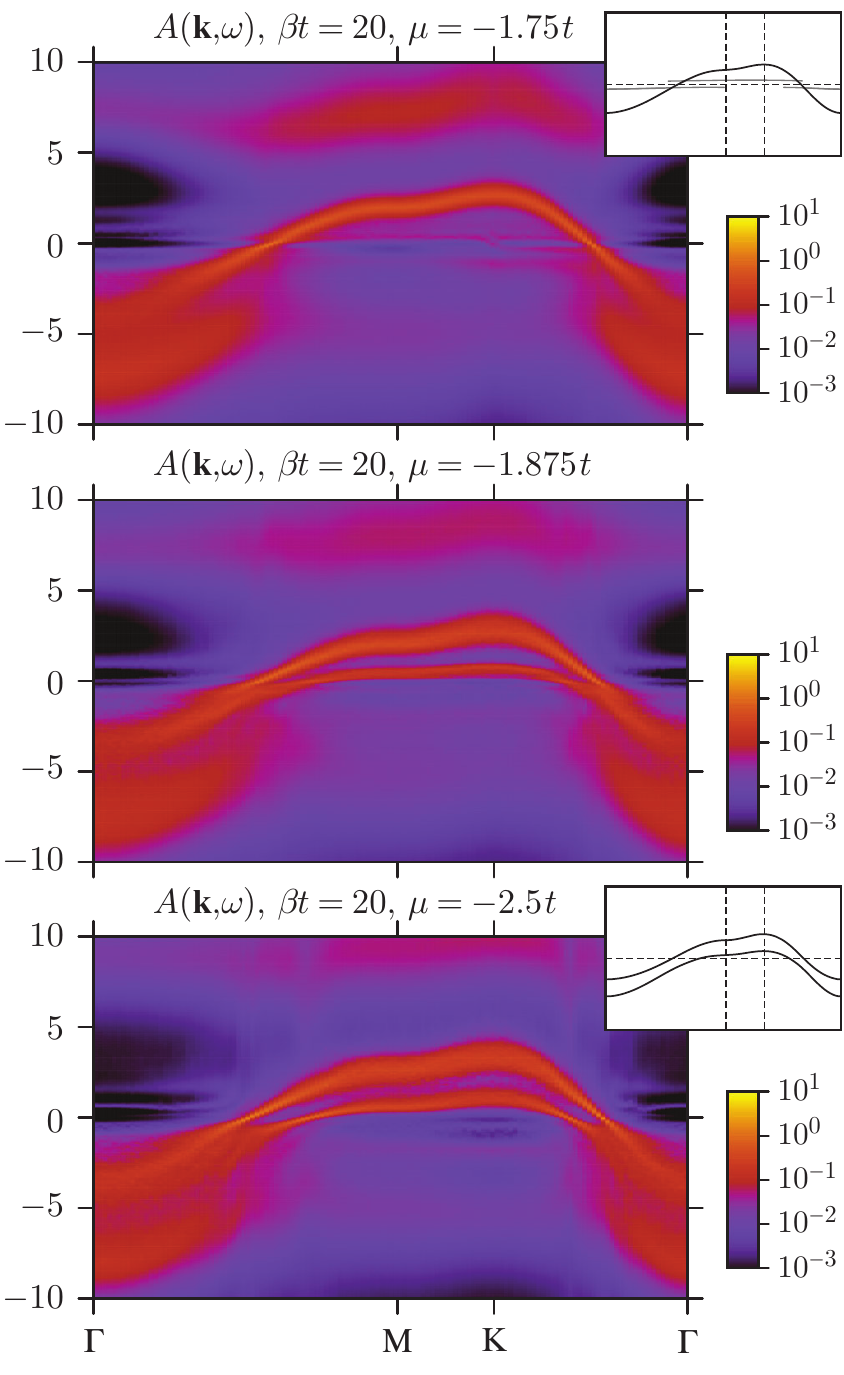}
\caption{\label{Akom.fig} 
The amplitude of the single particle spectral function,
as defined by Eq.~\eqref{Akom.eq}, is plotted for various values of the chemical potential. 
The lower inset shows the mean-field band structure consisting
of two quasiparticle bands of mixed $c$ and $f$ character. The upper inset shows the single $c$-only band completely
decoupled from the gapped, nearly flat band of the singlet-bound $f$ fermions.}
\end{figure}

The nature of the distinct metallic phases on either side of the transition is best 
understood in terms of the single particle spectral function, 
\begin{equation}
\label{Akom.eq}
A(\vec{k},\omega) = - \Imag \Tr G(\vec{k},\omega), 
\end{equation} 
plotted in Fig.~\ref{Akom.fig}.
For $\mu < \mu_c $, and as exemplified by the data set at $\mu = -2.5 t$, the low-energy 
coherent features of the spectral function compare favorably with a slave 
boson approximation leading to mass-renormalized, hybridized bands. This state has a Luttinger volume that includes both 
$f$ and $c$ fermions, and the band with the largest Fermi volume has dominant $f$ character. As a function of 
the chemical potential, the effective mass of the $f$ band grows; beyond $\mu_c$, the $f$ band drops 
out of the low-energy physics altogether. This can be understood at the \emph{static}
mean field level by a conventional slave boson theory in competition with local singlet 
formation in the first layer. The transition is signaled by the appearance of an 
anomalous expectation value $\Delta_{\vec{i}\vec{j}} \sim (t_f^2/U_f)\sum_{\sigma} \bracket{f_{\vec{i},\sigma}^\dagger f_{\vec{j},\sigma}}$. The upper inset of Fig.~\ref{Akom.fig} shows the band structure that results 
when this singlet order parameter has the configuration depicted in Fig.~\ref{Model.fig}(c).

We understand this transition to be of the band-selective Mott type,
in which a half-filled band with dominant $f$ and sub-dominant $c$ character is 
gapped beyond $\mu_c$. This interpretation is supported by the fact that, beyond $\mu_c$, 
$\bracket{\hat{n}_f}$ does not saturate to unity, as in the case of an orbital-selective 
Mott transition 
\footnote{The fact that $\bracket{\hat{n}_f}$ remains less than unity has been confirmed 
by calculations on smaller four-sites clusters, for which lower temperatures can be reached at $\mu > \mu_c$.}.
The data in Fig.~\ref{Akom.fig} show the typical Mott-Hubbard transfer of spectral weight from the 
upper band at $ \omega \sim 6 t$ ($\mu = -1.75t$) down to the Fermi energy as a function of 
decreasing chemical potential~\cite{Eskes91}. In the generic Mott-Hubbard scenario, doping occurs when the chemical potential reaches the lower Hubbard band, 
which in the present case is situated at roughly $\omega \sim -6t$ for $\mu = -1.75t$. In the band-selective 
Mott transition, doping is provided by changing the occupation of the ungapped band. 

\begin{figure}
\includegraphics{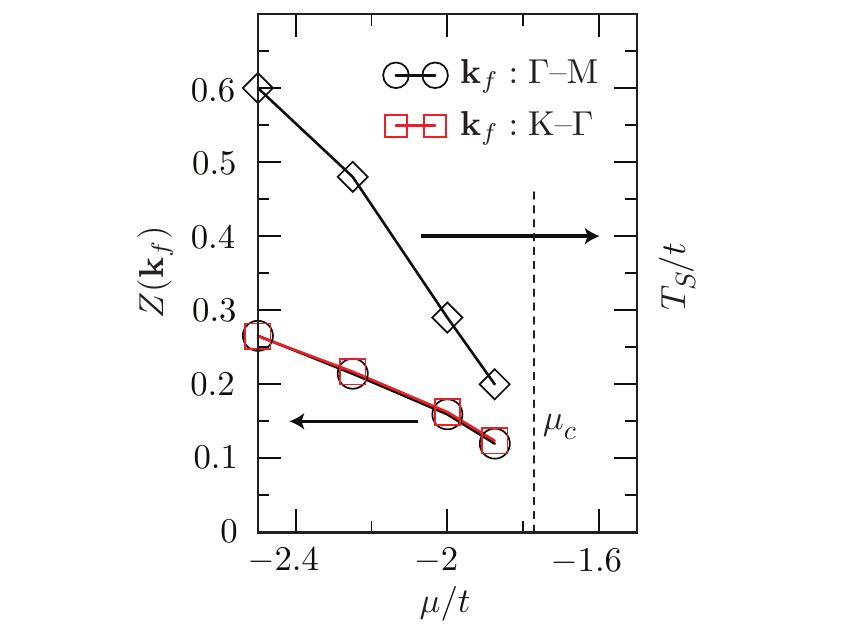} 
\caption{\label{mstar.fig} 
Quasiparticle residue as defined by Eq.~\eqref{Z.eq} and the spin scale $T_S$ as a function 
of chemical potential. The data is extracted from simulations at $N_c = 8$ and $\beta t = 20$.}
\end{figure}

To accurately estimate the effective mass renormalization at $\mu < \mu_c$, we consider the self-energy, 
$\Sigma_{f\!f}(\vec{k}, \omega_m)$, as defined by 
$G_{f\!f}^{-1}(\vec{k},i \omega_m) = i \omega_m - \varepsilon_f(\vec{k}) + \mu - \Sigma_{f\!f}(\vec{k}, \omega_m)$,
with the Green function taken from Eq.~\eqref{Greenk.eq}.
Since at $\mu < \mu_c$ the self-energy is dominated by its frequency dependence, the 
effective mass renormalization is very well approximated by the inverse quasiparticle residue, 
\begin{equation}
\label{Z.eq}
 \frac{m^{\star}}{m} \propto Z^{-1}(\vec{k}) = 1 - \left. \frac{\partial \Imag \Sigma_{f\!f}(\vec{k},i \omega_m)} 
 {\partial \omega_m} \right|_{\omega_m \rightarrow 0}.
\end{equation}
This quantity is plotted in Fig.~\ref{mstar.fig} for Fermi wave vectors of the $f$ band along the 
$\Gamma$--M and K--$\Gamma$ directions in the Brillouin zone. A monotonic increase
of the effective mass is observed, but its divergence is preempted by the first order transition 
at $\mu = \mu_c$. 

\begin{figure}
\includegraphics{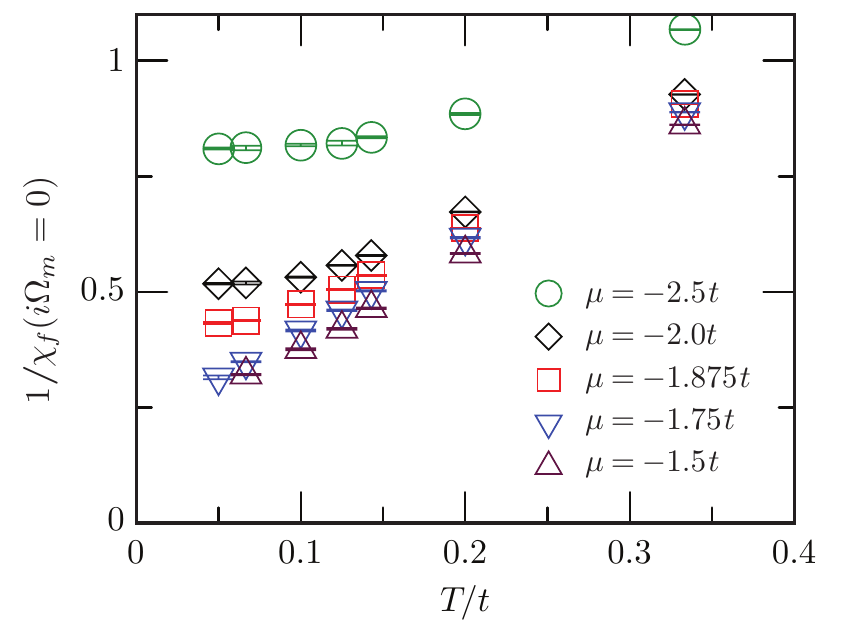}
\caption{\label{Spin_suecp.fig}
 Inverse local spin susceptibility as a function of temperature for the $N_c=8$ cluster.} 
\end{figure}

The growth of the effective mass corresponds to a decrease of the coherence temperature, 
$T_{\rm coh}$, below which Fermi liquid behavior manifests itself. To illustrate this, 
we have computed the local spin susceptibility on the cluster, as defined by
\begin{equation}
 \chi_f(i \Omega_m) = \frac{1}{L_c^2} \sum_{\vec{x}} \int_{0}^{\beta} \!{\rm d} \tau\,
 e^{i \Omega_m \tau} \langle \vec{S}_f(\vec{x},\tau) \cdot \vec{S}_f(\vec{x}) \rangle.
\end{equation}
Below the coherence temperature, $\chi_f(i \Omega_m\!=\!0)$ is expected to be 
temperature independent. On the other hand, in the temperature region 
$ T_{\rm coh} < T \ll U$, it should exhibit Curie-Weiss behavior---the signature
of local moment formation.
Precisely this behavior is seen in Fig.~\ref{Spin_suecp.fig}. As the chemical potential grows
from $\mu = -2.5t$ to $\mu = -1.875t$, the crossover temperature scale between the Curie-Weiss-like
and temperature-independent $\chi_f(i \Omega_m\!=\!0)$ tracks the decrease of the
inverse effective mass and coherence temperature. Beyond the phase transition, 
$\mu =-1.75t > \mu_c$, only Curie-Weiss behavior is apparent in the considered temperature range. 

\begin{figure}
\includegraphics{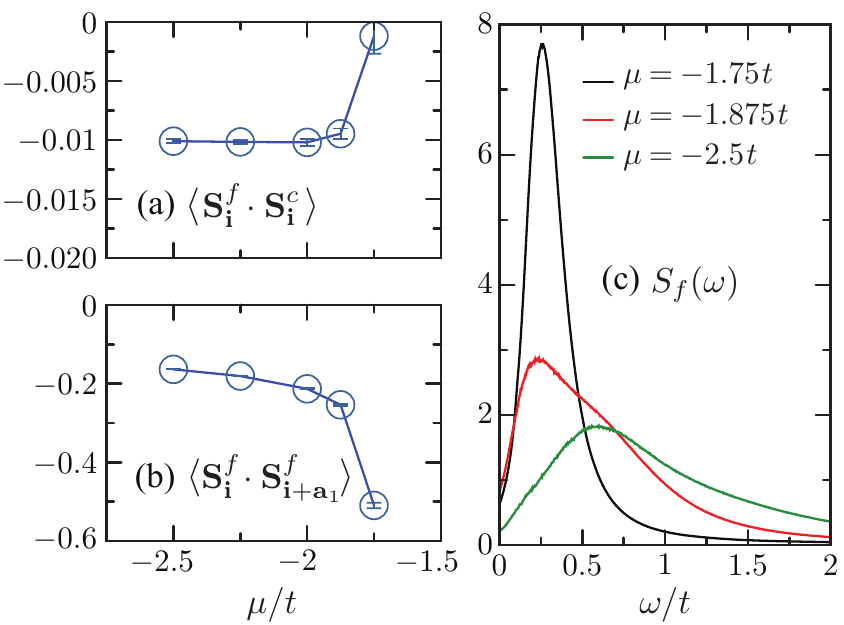} 
\caption{\label{Spin_eq.fig}
 (a) Intracell spin-spin correlations between $c$ and $f$ fermions. 
 (b) Nearest-neighbor spin-spin correlations between $f$ fermions. 
 (c) Local dynamical spin structure factor.} 
\end{figure}

Fig.~\ref{Spin_eq.fig} shows that the band-selective Mott transition is 
linked to a {\it sudden} growth of the antiferromagnetic correlations between nearest-neighbor
$f$ fermions and to a decrease in the intracell $c$-$f$ spin-spin correlations.
This data supports the picture that, in the band-selective Mott insulating phase, the 
$f$ quasiparticles are bound into spin singlets amongst themselves. 
The {\it gapping} of the spin and charge degrees of freedom of the $f$ quasiparticles at $\mu > \mu_c$ allows 
for the decoupling of $f$ and $c$ quasiparticles: a $c$ quasiparticle at the Fermi level cannot scatter 
off an $f$ quasiparticle due to the absence of phase space. 

The local dynamical spin structure factor, 
\begin{equation}
 S_f(\omega) = \Imag \frac{\chi_f(\omega)}{1-e^{-\beta \omega}},
\end{equation}
(see Fig.~\ref{Spin_eq.fig}) shows a depletion of spectral weight at low energies on both 
sides of the transition and a considerable sharpening of the line shape in the band-selective 
Mott insulating state. At $\mu < \mu_c$, we can interpret the data within an itinerant fermion picture
where the mass enhancement prior to the band-selective Mott transition is taken into account 
by a renormalization of the hybridization $V$ and hopping $t_f$ as in a slave boson approach \cite{Benlagra08}.
Following this modeling, the peak position in $S_f(\omega) $, which we will denote by $T_S$, 
is expected to track the coherence temperature or, equivalently, the inverse effective mass. 
The quantity $T_S$ is plotted in 
Fig.~\ref{mstar.fig} alongside $Z(\vec{k}_f)$ and confirms the above expectations to a good degree. 
At $\mu > \mu_c$, $S_f(\omega)$ should be interpreted within a localized $f$ fermion picture, in which case 
the peak position is a measure of the excitation energy required to break the singlet state of the $f$ fermions. 

We can summarize our results using the terminology of heavy fermions \cite{Loehneysen_rev}. 
The band-selective Mott transition corresponds to a Kondo breakdown in which the $f$ fermions 
drop out of the Luttinger volume. Within our model, it appears that this transition is first order: the 
reduction of the coherence temperature and the enhancement of the effective mass
is interrupted by the formation of a \emph{spin gapped} Mott insulating state of the $f$ fermions. The nature of this 
band-selective Mott insulating state is very dependent on the cluster topology. In this work, 
we have considered only clusters with an even number of unit cells---thereby implicitly allowing for 
spin gapped insulating states of the $f$ fermions within the CDMFT calculation. Despite
the breaking of translation invariance inherent to the CDMFT, the Luttinger sum rule still holds when 
formulated in the Brillouin zone of the supercell Bravais lattice. We can only speculate as to the nature 
of this state when the cluster size diverges, but we cannot exclude the intriguing possibility that it 
smoothly connects to fractionalized Fermi liquids \cite{Senthil03,Senthil04} with no lattice 
and spin symmetry breaking but a Luttinger volume encompassing only the $c$ fermions. 

\noindent 
\emph{Acknowledgments}. The numerical calculations were carried out at the LRZ-M\"unich and
the J\"ulich Supercomputing center. We thank those institutions for their generous allocation of CPU time. 
KSDB thanks the Humboldt foundation for financial support as well as the FFA and DFG 
under grant number AS120/4-2.

\end{document}